\ifx\pdfoutput\undefined
  \documentclass[prb,twocolumn,superscriptaddress,dvipdfmx]{revtex4-2}
\else
  \pdfoutput=1
  \documentclass[prb,twocolumn,superscriptaddress]{revtex4-2}
\fi

\usepackage{graphicx}
\usepackage{dcolumn}
\usepackage{bm}
\usepackage[normalem]{ulem}
\usepackage{booktabs}
\usepackage{xcolor}
\usepackage[colorlinks=true, linkcolor=blue, citecolor=blue, urlcolor=blue]{hyperref}

\begin{document}

\preprint{APS/123-QED}

\title{
Theoretical exploration of ambient-pressure superconductivity in the bilayer nickelate: 
A strategy of atomic substitution for tuning chemical pressure
} 

\author{Tomo-Hiko Hamada}
\email{m25j3049b@edu.tottori-u.ac.jp}
\affiliation{Faculty of Engineering, Tottori University, 4-101 Koyama-cho Minami, Tottori, Tottori 680-8552, Japan}
\author{Kensei Ushio}
\affiliation{Faculty of Engineering, Tottori University, 4-101 Koyama-cho Minami, Tottori, Tottori 680-8552, Japan}
\author{Shu Kamiyama}
\affiliation{Department of Physics, University of Osaka, 1-1 Machikaneyama-cho, Toyonaka, Osaka 560-0043, Japan}
\author{Kazuhiko Kuroki}
\affiliation{Department of Physics, University of Osaka, 1-1 Machikaneyama-cho, Toyonaka, Osaka 560-0043, Japan}
\author{Hirofumi Sakakibara}
\email{sakakibara@tottori-u.ac.jp}
\affiliation{Faculty of Engineering, Tottori University, 4-101 Koyama-cho Minami, Tottori, Tottori 680-8552, Japan}
\affiliation{Advanced Mechanical and Electronic System Research Center(AMES), Tottori University, 4-101 Koyama-cho Minami, Tottori, Tottori 680-8552, Japan}

\date{\today}

\begin{abstract}
The possibility of ambient-pressure superconductivity in $\mathrm{La_3Ni_2O_7}$ is theoretically explored via chemical substitution. Using first-principles and many-body calculations, we propose substituting $\mathrm{Ba}$ at the lanthanide site to exert chemical pressure, thereby mitigating the octahedral distortions that suppress superconductivity. Phonon calculations demonstrate the dynamical stability of the high-symmetry tetragonal phase in $\mathrm{La_{3-\it{x}}Ba_{\it{x}}Ni_2O_7}$ at a small doping ratio of $x \sim 0.15$. By applying the fluctuation exchange approximation to an effective model Hamiltonian derived from this stable structure, we calculate the Eliashberg eigenvalue $\lambda$ to evaluate the relative superconducting transition temperature $T_\mathrm{c}$ to existing unconventional superconductors. We conclude that $s_\pm$-wave superconductivity is likely to emerge, as the required $\mathrm{Ba}$ concentration ($x \sim 0.15$) is sufficiently small to minimize the detrimental suppression of superconductivity induced by hole doping.
\end{abstract}

\pacs{74.20.Mn,74.70.−b}
\maketitle
\section{Introduction}
The discovery of high-temperature superconductivity with a $T_\mathrm{c}$ near 80--90 K in $\mathrm{La_3Ni_2O_7}$~\cite{MWang,MWangReview} in 2023 has stimulated striking interest for bilayer nickelates. 
In the bulk system, superconductivity emerges only after a structural phase transition from the orthorhombic to the tetragonal phase under applied pressures of 10 to 20 GPa. Since 2025, ambient-pressure superconductivity has been realized in thin-film samples of bilayer nickelates~\cite{Ko2025,QKXuePrNature,HwangPrNature,QKXueThin}. These advances have enabled the observation of the superconducting gap via ARPES~\cite{QKXue2502.17831} and STM/STS~\cite{HHWen2506.01788}, revealing distinction from pure $d$-wave pairing.

From a theoretical viewpoint, the essence of superconductivity in $\mathrm{La_3Ni_2O_7}$ can be effectively captured by the bilayer Hubbard model. The potential for high-$T_\mathrm{c}$ $s_\pm$-wave superconductivity in a simple bilayer Hubbard model with a large interlayer hopping $t_\perp$ was originally proposed in the early 2000s~\cite{KA,MaierScalapino}. Building upon these studies, Nakata {\it et al.}~\cite{Nakata} performed first-principles calculations in 2017 to propose $\mathrm{La_3Ni_2O_7}$ as a candidate superconductor characterized by a nearly half-filled bilayer Hubbard model formed by the $d_{3z^2-r^2}$ orbital (hereafter abbreviated as $d_{z^2}$). Currently, both single-orbital~\cite{Nomurabilayer,DMRG-Kaneko} and multi-orbital bilayer Hubbard model, incorporating the $d_{x^2-y^2}$ orbital in addition to the $d_{z^2}$ orbital, are being extensively studied in connection with $\mathrm{La_3Ni_2O_7}$~\cite{LuoPRL,Yang,sakakibarala327,ZhangDagotto,HanghuiChen,PhysRevB.109.104508,InoueOnari,2503.18877,Eremindwave}. Furthermore, various $t\text{--}J$-type bilayer models have been broadly investigated as effective multi-orbital descriptions~\cite{Luo2024,PhysRevB.108.174511,LuoDXYao_npjqm,PhysRevB.110.024514}.
Related to them, multi-orbital ladder model is also being studied in density matrix renormalization group calculation~\cite{DMRG-Kaneko,DMRG-Kakoi}.

Numerous theoretical studies have proposed various scenarios regarding the nature of this superconductivity. For example, we previously demonstrated that the spin-fluctuation mechanism, treated within the fluctuation exchange approximation (FLEX), can explain the relative strength of superconductivity among unconventional superconductors (see the Figure in the Appendix of Ref.~\cite{Ushio}). In that study, we concluded that finite-energy spin fluctuations render the $s_\pm$-wave pairing symmetry insensitive to the presence or absence of the $\gamma$ pocket~\cite{PengARPES,2504.16372}. This conclusion was promptly re-examined and reinforced using the variational Monte Carlo (VMC) method~\cite{WatanabeVMC}, enable to treat strong electronic correlations more accurately. Furthermore, this scenario is consistent with theoretical investigations reported by other groups~\cite{RyeeDMFT,RyeeRPA,Gao2502.19840}.

However, a fundamental question remains: what is the minimum prerequisite for superconductivity? Previously, based on first-principles calculations and FLEX, we theoretically predicted a pressure-induced structural transition in $\mathrm{La_4Ni_3O_{10}}$, a trilayer counterpart of $\mathrm{La_3Ni_2O_7}$, from monoclinic to tetragonal symmetry, as well as the emergence of superconductivity in the high-pressure tetragonal phase~\cite{sakakibara4310}. A signature of superconductivity was first reported in Ref.~\cite{sakakibara4310} via resistivity measurements, which was subsequently reinforced by independent observations~\cite{La4Ni3O10Nature,PhysRevX.15.021005,NagataLa4310,La4310Chin}. These findings suggest that the tetragonal $I4/mmm$ space group is necessary for the realization of superconductivity.
In several experiments, spin/charge density waves or charge ordering have been observed in both $\mathrm{La_3Ni_2O_7}$ and $\mathrm{La_4Ni_3O_{10}}$ at ambient pressure~\cite{KakoiNMR,YashimaNQR,LeeMukuda}, 
which appear to be intimately linked to octahedral deformations and/or orthorhombicity. These observations lead to the speculation that such orderings compete with or suppress superconductivity under ambient pressure. Moreover, recent structural analysis indicates that the space group of the low pressure orthorhombic phase is $Am2m$ rather than $Amam$ as previously believed, and there is a charge disproportionation in this phase, which would be unfavorable for superconductivity\cite{MisawaArima,SasakiArima}. In thin films of $\mathrm{La_3Ni_2O_7}$ and $(\mathrm{La,Pr})_3\mathrm{Ni_2O_7}$ exhibiting ambient-pressure superconductivity~\cite{Ko2025,QKXuePrNature,HwangPrNature,QKXueThin}, structural analyses indeed confirm an $I4/mmm$ space group (or one very close to it)~\cite{QKXuePrNature}. It is widely believed that compressive strain from the substrate, which possesses a smaller lattice constant, acts analogously to hydrostatic pressure in bulk systems~\cite{HirschfeldThin,BotanaThinPRB,Ushio,2502.01624,2501.08204,85qv-ncxb,DXYao_2503.17223,2501.14665,Rhods_tetra,Shao9t6n-jqr5}. On the other hand, a recent experimental study suggests the absence of superconductivity even within the $I4/mmm$ symmetry~\cite{Shi2025}. Thus, whether and how structural deformations govern superconductivity remains a subject of intense debate.

In addition to the physical pressure effect, partial chemical substitution at the lanthanum 
site has become a hot topic. For instance, Pr and Nd substitutions have led to a significant enhancement of the superconducting volume fraction~\cite{LaPr327} as well as higher $T_\mathrm{c}$~\cite{LaNd100K,Sm100K}. It is believed that substituting a lanthanide with a smaller ionic radius increases the interlayer hopping $t_\perp$, thereby elevating $T_\mathrm{c}$. Meanwhile, the impact of carrier doping is also being intensively investigated~\cite{KrienerNd,OxygendopingPRB,OzoneLa327}. The negative impact of hole doping on superconductivity was predicted shortly after the discovery of $\mathrm{La_3Ni_2O_7}$~\cite{sakakibarala327} and subsequently demonstrated in thin-film samples~\cite{LaSr_Nie}. The physical origin of this suppression is the weakening of spin fluctuations in the $d_{z^2}$ orbital as it shifts away from half-filling~\cite{sakakibarala327}. 

Even after the discovery of ambient-pressure superconductivity in thin-film bilayer nickelates~\cite{Ko2025,QKXuePrNature,HwangPrNature,QKXueThin}, there remains a strong motivation to achieve ambient-pressure superconductivity in bulk systems, which may be achieved in the quintuple-layer counterpart~\cite{Bulk5layer} of the infinite layer nickelates, as this would enable the measurement of thermodynamic bulk quantities. Based on the above discussions, an effective strategy for realizing superconductivity is to stabilize high-symmetry crystal structures~\cite{MWang,JACSinsituXRD}, which are typically accessed via external hydrostatic pressure~\cite{MWang}. Several theorists, including ourselves, have explored an alternative route using chemical pressure. For example, actinium-containing compounds have been proposed as candidates that adopt a tetragonal structure at ambient pressure~\cite{2403.11713,PhysRevMaterials.8.044801,commentact}. We previously proposed a novel material, $\mathrm{Sr_3Ni_2O_5Cl_2}$~\cite{Sr3252}, taking advantage of the large ionic radii of $\mathrm{Sr}$ and $\mathrm{Cl}$. Following this proposal, the tetragonal $I4/mmm$ symmetry at ambient pressure was experimentally confirmed by Yamane {\it et al.}~\cite{Yamane3252}, although superconductivity has not yet been observed (likely due to lattice defects associated with $\mathrm{Cl}$ bonding).

\begin{figure}
\hspace*{-0.5cm}
	\includegraphics[width=7.5cm]{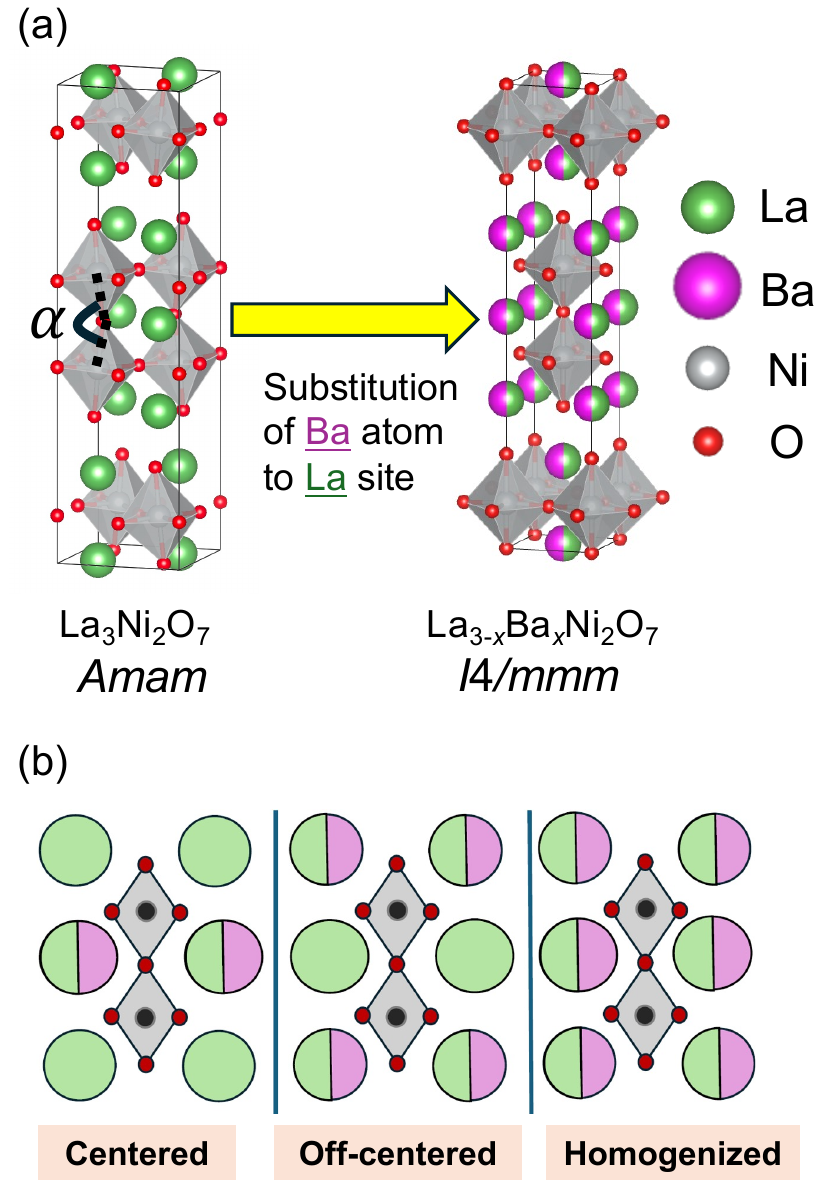}	
	\caption{(a) Schematic picture of the chemical pressure effect on the octahedral distortion and angle $\alpha$,
	and (b) the pattern of Ba distribution adopted in the present VCA calculation.
	Here, the crystal structure is drawn by VESTA~\cite{VESTA}.
	}
	\label{fig1}
\end{figure}

In this paper, we propose a strategy to achieve ambient-pressure superconductivity in $\mathrm{La_3Ni_2O_7}$ through partial elemental substitution. A material where the lanthanum site is partially replaced by $\mathrm{Sr}$ with larger ionic radius has already been synthesized, and this material indeed exhibits $I4/mmm$ tetragonal phase at ambient pressure~\cite{HHWen2026}. However, in this case, large number of holes are doped in the $d_{z^2}$ orbitals, which would be destructive for superconductivity, as mentioned above. Here, we examine the impact of $\mathrm{Ba}$ substitution at the  lanthanum site on the crystal structure and the degree of octahedral distortion. The idea is that smaller amount of substitution may be enough to attain tetragonal symmetry because $\mathrm{Ba^{2+}}$ is larger than $\mathrm{Sr^{2+}}$. 
Figure~\ref{fig1}(a) schematically illustrates the overall concept of this study.
By estimating the bonding angle $\alpha$ between two point-sharing $\mathrm{NiO_6}$ octahedra along $c$ axis, we demonstrate that a straight bond angle of $\alpha (= 180^\circ)$ is achieved even at a small $\mathrm{Ba}$ substitution ratio of $x \sim 0.15$--$0.20$ in $\mathrm{La_{3-\it{x}}Ba_{\it x}Ni_2O_7}$. Furthermore, phonon calculations confirm the dynamical stability of the tetragonal $I4/mmm$ structure. Applying FLEX calculations to the two-orbital bilayer model derived from this stable $I4/mmm$ structure, we evaluate the relative $T_\mathrm{c}$ in comparison with existing unconventional superconductors. These results lead us to conclude that our substitution strategy is highly promising for achieving ambient-pressure superconductivity.

\section{Method}

To calculate the bond angle $\alpha$, we perform first-principles structural optimization assuming the $Amam$ space-group symmetry, which was experimentally proposed for the bulk system at ambient pressure~\cite{MWang}. All calculations are carried out using the PBEsol exchange-correlation functional~\cite{PBEsol} and the projector augmented-wave (PAW) method as implemented in the Vienna {\it ab initio} Simulation Package (VASP)~\cite{vasp1,vasp2,vasp3,vasp4}. The core-electron configurations in the PAW potentials are set to $[\mathrm{Kr}]4d^{10}$, $[\mathrm{Ar}]$, and $[\mathrm{He}]$ for $\mathrm{La(Ba)}$, $\mathrm{Ni}$, and $\mathrm{O}$, respectively. A plane-wave cutoff energy of 550 eV is employed for the Kohn-Sham orbitals. For Brillouin-zone integration, we use an $8 \times 8 \times 8$ ${\bm k}$-mesh~\cite{kconv} for the $Amam$ structure, with Gaussian smearing applied at an energy width of 0.1 eV. Structural relaxation is performed until the Hellmann-Feynman forces on all atoms fall below 0.001 eV \AA$^{-1}$. The Virtual Crystal Approximation (VCA) is adopted to model the effect of partial $\mathrm{Ba}$ substitution at the lanthanide site. Hereafter, $x_\mathrm{c}$ denotes the critical doping ratio corresponding to the structural transition between the $Amam$ and $I4/mmm$ phases.

For the case of $\alpha = 180^\circ$, we performed phonon calculations to examine the dynamical stability of the $I4/mmm$ structure. To obtain the fully optimized geometry of the $I4/mmm$ phase, a $12 \times 12 \times 12$ ${\bm k}$-mesh was employed. Phonon frequencies were calculated using the finite displacement method as implemented in the {\textsc{Phonopy}} package~\cite{phonopy,phonopy1,phonopy2}. These supercell calculations were carried out using a $2 \times 2 \times 2$ supercell accompanied by a $6 \times 6 \times 6$ ${\bm k}$-mesh. The absence of imaginary phonon modes across the Brillouin zone confirms the dynamical stability of the $I4/mmm$ phase.

To analyze the superconducting properties, we construct model Hamiltonians based on the crystal structures determined by the aforementioned procedure. For the band-structure calculations, we incorporate the $+U$ correction within Dudarev's formulation~\cite{GGAU} using $U = 3\text{ eV}$. 
Maximally localized Wannier functions~\cite{Marzari,Souza} are then constructed using the {\textsc{Wannier90}} software package~\cite{Wannier90}, from which the hopping parameters among the Wannier orbitals are extracted. Specifically, we construct a two-orbital bilayer tight-binding model using the $\mathrm{Ni}\text{--}3d_{x^2-y^2}$ and $3d_{3z^2-r^2}$ orbitals as initial projections. For this step, a $12 \times 12 \times 12$ ${\bm k}$-mesh is employed for the primitive unit cell of the $I4/mmm$ structure, maintaining consistency between the first-principles calculations and the Wannier fitting procedure. Subsequently, the two-orbital bilayer Hubbard model is obtained by adding the on-site interaction terms to the tight-binding Hamiltonian. We employ the constrained random phase approximation (cRPA)~\cite{cRPA} to quantitatively determine the on-site Coulomb ($U, U'$) and exchange interaction ($J, J'$) parameters for the model.


We take into account the electron correlation effects for the two-orbital bilayer Hubbard model within FLEX approximation~\cite{Bickers,Bickers1991} as was done in Ref.~\cite{sakakibarala327}. We calculate the self-energy induced by the spin-fluctuation formulated as shown in the literatures~\cite{Lichtenstein,mFLEX1,mFLEX2}, in a self-consistent calculation.
The formulation of FLEX is given in the APPENDIX.
The real part of the self-energy at the lowest Matsubara frequency is subtracted in the same manner with Ref.~\cite{Ikeda_omega0}
to maintain the band structure around the Fermi level obtained by first-principles calculation.
The obtained Green function and the pairing interaction, mediated mainly by spin fluctuations, are plugged into the linearized Eliashberg equation.
Since the eigenvalue $\lambda$ of the linearized Eliashberg equation reaches unity at $T=T_c$, 
we adopt it as a measure of superconductivity at a fixed temperature, $T=0.01$ eV. 
The eigenfunction of the linearized Eliashberg equation corresponds to the anomalous self energy in the original (non-linearized) Eliashberg equation, but for convenience, we will call the eigenfunction (with the largest eigenvalue)  at the lowest Matsubara frequency $i\omega$(=$i\pi k_{\rm B}T$) the ``superconducting gap function''. 
In this work, the gap functions are calculated in the orbital basis and subsequently transformed into the band basis using the unitary matrix derived from the diagonalization of the unperturbed normal-state Hamiltonian.
We take a 16$\times$16$\times$4 ${\bm k}$-point mesh and 2048 Matsubara frequencies for $I4/mmm$.
For all cases of the present calculation, we obtain the eigenfunction of $s\pm$-wave superconductivity.
Note that the pairing interaction kernel in this equation was obtained from the FLEX Green function as a purely electronic one (i.e. phonon-mediated pairing interaction was not considered), which is mainly dominated by spin fluctuations in the present case.

\section{Results}

\subsection{Impact of Ba substitution on crystal structure}

\begin{figure}
\hspace*{-0.5cm}
	\includegraphics[width=7.5cm]{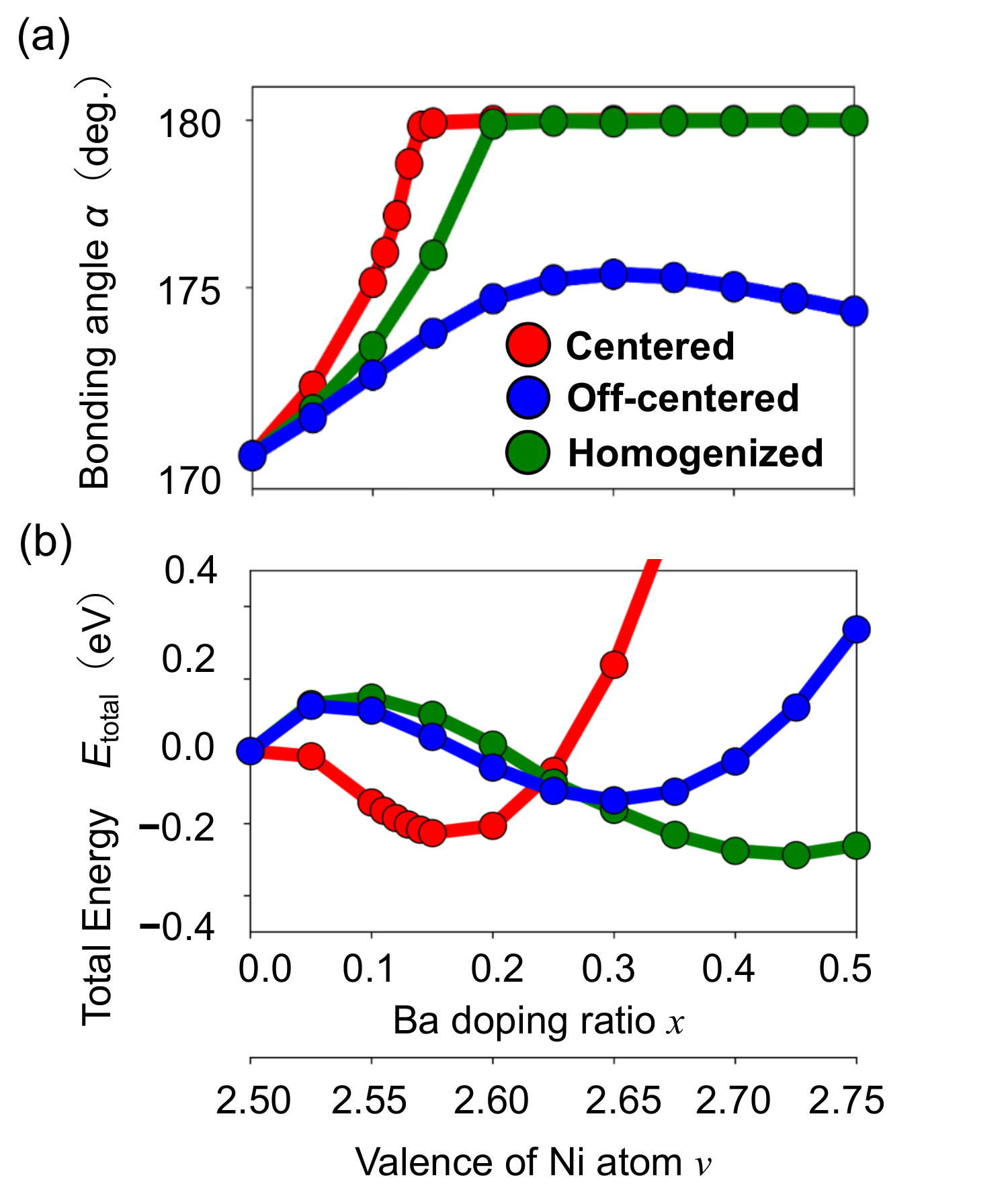}	
	\caption{ (a)The bonding angle $\alpha$ and the relative total energy $E_{\mathrm{total}}$ for the three distinct cases described in Figs.~\ref{fig2} (a)-(c) are plotted as a function of the substitution ratio $x$, where the reference point for $E_{\rm total}$ is the state of $x=0$.
	Here, the corresponding valence $v$ of the Ni atom is also indicated.
	The pattern of Ba distribution is defined in Fig.~\ref{fig1}(b).
	}
	\label{fig2}
\end{figure}

To estimate the realistic spatial distribution of substituted $\mathrm{Ba}$ atoms, we compare the total energies $E_{\rm total}$ of the three configuration cases illustrated in Fig.~\ref{fig1}(b): (i) $\mathrm{Ba}$ localizes selectively at the inner lanthanide site (Centered), (ii) $\mathrm{Ba}$ resides preferentially at the outer lanthanide site (Off-centered), and (iii) $\mathrm{Ba}$ is distributed homogeneously across all lanthanide sites (Homogenized). 
The corresponding results are summarized in Figs.~\ref{fig2}(a) and \ref{fig2}(b). 
Figure~\ref{fig2}(a) shows that the bond angle $\alpha$ reaches $180^\circ$ at $x \sim 0.15$ in the Centered case, which exhibits the fastest evolution among the all cases.
Even in the Homogenized case, $\alpha$ reaches $180^\circ$ at $x \sim 0.20$, implying that the critical doping ratio $x_c\sim 0.15$--$0.20$.
From the panel (b) of Fig.~\ref{fig2}, we find that the Centered configuration is the most energetically stable at low doping levels, whereas the Homogenized configuration becomes the most stable at higher substitution ratios. 
The Off-centered configuration is the least energetically favorable over the entire doping range except $x\sim 0.25$, where the all cases have the similar amount of energy.
In a realistic system, we therefore can expect $\mathrm{Ba}$ atoms to dominantly occupy the inner site at small $x$, with the occupancy at outer sites gradually increasing for larger $x$. 
In the following, we compare the physical properties between the Centered and Homogenized cases.

\subsection{Band dispersions of phonon and electron}
\begin{figure*}
\hspace*{-0.5cm}
	\includegraphics[width=15cm]{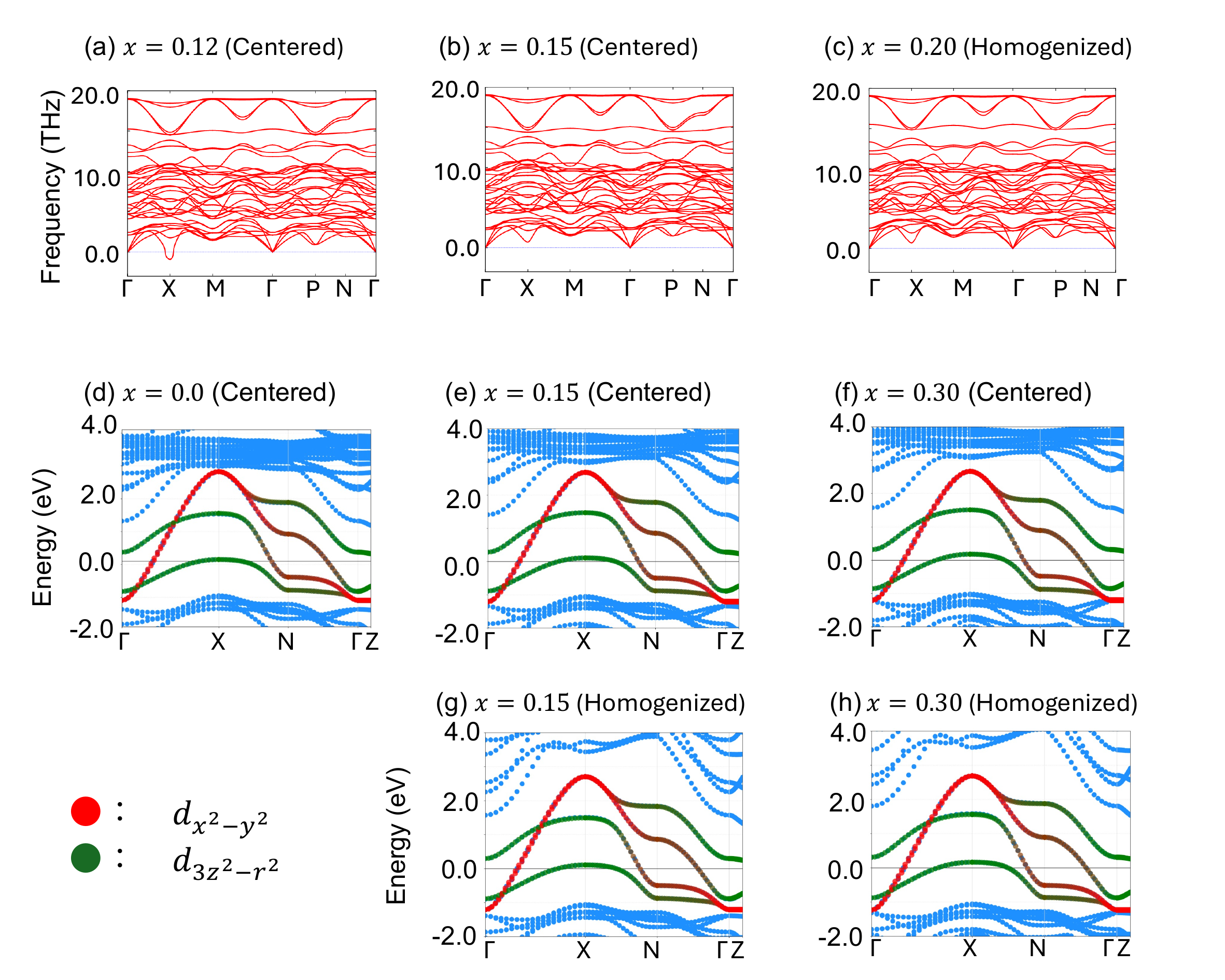}	
	\caption{Band dispersions of (a)-(c) phonon and (d)-(h) electron (relative to the Fermi level) are displayed.
	Assumed Ba-distribution type and doping ratio are indicated in each panel.
	In the panels (d)-(h), the orbital weights are depicted by the width of color coded lines.
	}
	\label{fig3}
\end{figure*}

Figures~\ref{fig3}(a) and (b) show the phonon dispersions for $x = 0.12$, $x = 0.15$, respectively. 
The absence of imaginary phonon modes demonstrates that the $I4/mmm$ structure is dynamically stable for the case of panels (b) but unstable for (a). 
Assuming that all $\mathrm{Ba}$ atoms occupy the inner site, this indicates that the critical doping concentration lies in the range $0.12 < x_\mathrm{c} < 0.15$. In reality, a non-negligible fraction of $\mathrm{Ba}$ atoms occupies the outer site.
Therefore $x_\mathrm{c}$ is expected to be slightly higher than idealized (Centered) case, though still lower than $x_\mathrm{c} \sim 0.20$ corresponding to the fully homogenized case. 
For the case, we have also confirmed the absence of imaginary modes as shown in panel (c). 

 Figures~\ref{fig3}(d)-(h) present the electronic band structures along with their orbital character. Across the entire doping range, the band dispersion exhibits an almost rigid-band-like shift. In addition, the difference between Centered and Homogenized cases are quite small at each doping ratio.

\subsection{Superconducting strength under Ba doping}

\begin{figure}
\hspace*{-0.5cm}
	\includegraphics[width=9cm]{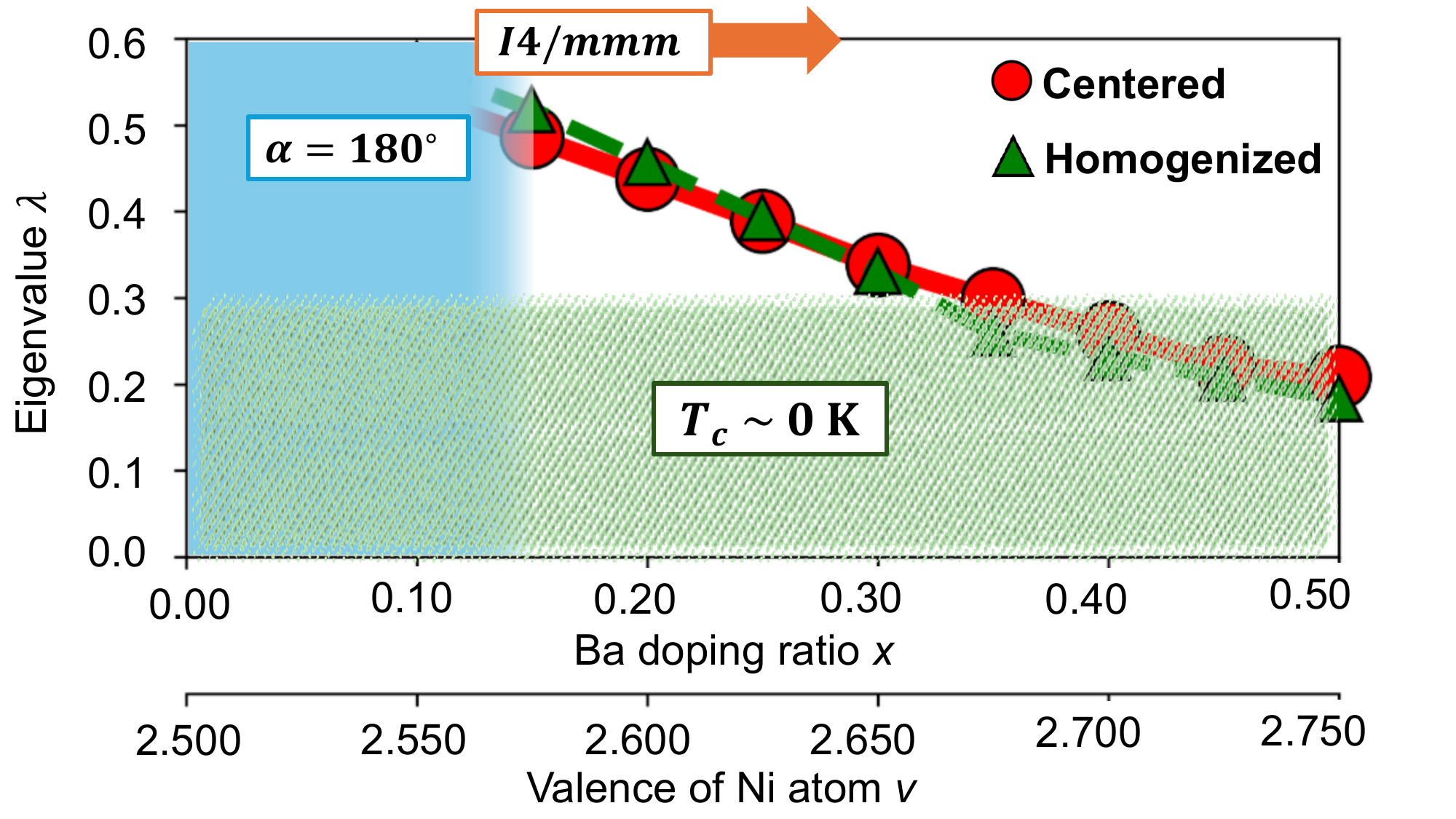}	
	\caption{The eigenvalues $\lambda$ of the Eliashberg equation are plotted against the $\mathrm{Ba}$ doping ratio $x$. As discussed in the main text, superconductivity is likely suppressed in the $x < 0.15$ region due to competing orders enhanced by octahedral distortions and/or orthorhombicity. Based on a comparison with existing unconventional superconductors, the emergence of superconductivity is unlikely for $\lambda < 0.3$ (see text).
	}
	\label{fig4}
\end{figure}

Here we address the superconducting properties in the presence of $\mathrm{Ba}$ doping. 
Figure~\ref{fig4} summarizes the overall results. 
As mentioned above, since the structural transition to the $I4/mmm$ phase is realized at $x = 0.15$ even in the idealized (Centered) configuration, the underdoped regime ($x < 0.15$) is expected to be dominated by competing order, such as spin/charge density waves~\cite{KakoiNMR,YashimaNQR,LeeMukuda} or charge ordering. 
In the region of $x \sim 0.15\text{--}0.20$ for both the Centered and Homogenized cases, the eigenvalue $\lambda$ of the Eliashberg equation is comparably large to those of existing unconventional superconductors, as shown in the Appendix of Ref.~\cite{Ushio}.
This result indicates that superconductivity emerge around this doping level. 
As $x$ increases further, $\lambda$ systematically decreases. This reduction originates from the negative impact of hole doping, which weakens spin fluctuations associated with the $d_{z^2}$ orbital away from half-filling, as described in our previous work~\cite{sakakibarala327}. Based on a simple extrapolation against the empirical trend in Ref.~\cite{Ushio}, regions with $\lambda < 0.3$ are expected to be non-superconducting; notably, this threshold is reached only at a relatively high doping level ($x \sim 0.3$).

\section{Discussion}

In this study, we constructed our model under the assumption that stabilizing the undistorted tetragonal phase ($\alpha = 180^\circ$) is a primary prerequisite for the emergence of superconductivity. However, it should be emphasized that our present calculations do not directly incorporate or simulate broken-symmetry phases that compete with superconductivity, such as spin/charge density waves~\cite{KakoiNMR,YashimaNQR,LeeMukuda} or charge ordering. Consequently, whether these competing orders are fully suppressed around $x \sim 0.15$ remains an open question, and a rigorous theoretical analysis of the interplay between structural distortions and competing order parameters is left for future work.

Another simplification in our approach lies in the treatment of chemical substitution. Although VCA effectively captures the spatial-average structural and electronic modulations induced by $\mathrm{Ba}$ substitution, the microscopic effects of random disorder--such as localized lattice distortions and local potential fluctuations--were not explicitly evaluated. Incorporating explicit spatial disorder (e.g., via large supercell calculations or real-space methods) requires formidable computational costs that exceed the scope of the present study, rendering it another important avenue for future investigation.

On the experimental side, while extensive studies have been conducted on $\mathrm{Sr}$-substituted bilayer nickelates, reports on $\mathrm{Ba}$-substituted systems remain absent to date. In thin-film geometry, a recent attempt to synthesize $\mathrm{Ba}$-substituted samples was unsuccessful due to synthetic challenges~\cite{BozovicBa}. Nevertheless, the synthesis conditions in bulk systems differ substantially from those in thin films, particularly with regard to strain states and thermal equilibrium during growth. We anticipate that bulk synthesis may offer a viable route to realize $\mathrm{Ba}$-substituted $\mathrm{La_3Ni_2O_7}$, thereby providing an experimental platform to test our theoretical predictions.


\section{Summary}

In summary, we have theoretically investigated the effects of $\mathrm{Ba}$ substitution in $\mathrm{La_{3-\it{x}}Ba_{\it{x}}Ni_2O_7}$ using first-principles calculations with VCA. 
Our total-energy calculations for evaluating site-selectivity reveal that at low doping levels, $\mathrm{Ba}$ atoms preferentially occupy the inner lanthanide site, whereas with increasing doping concentration, a uniform distribution across all lanthanide sites becomes thermodynamically more favorable. 
Based on this structural modeling, we demonstrate that the $I4/mmm$($\alpha = 180^\circ$) structure can be realized under ambient pressure at a small doping level of $x \sim 0.15$--$0.20$ (in other words, critical doping ratio for the structural transition is estimated to be $x_c\sim0.15$--$0.20$). Phonon calculations further confirm that this $I4/mmm$ structure at $x \sim 0.15$ is dynamically stable. 
To evaluate the superconducting properties under ambient pressure, we have construct an effective four-orbital model via construction of Wannier function based on GGA+$U$ calculation, with the on-site electron interactions determined by cRPA calculations. 
From a direct comparison with the benchmark reported in Ref.~\cite{Ushio} of the eigenvalue $\lambda$ obtained in FLEX,
we conclude that a strong superconducting instability is theoretically expected at $x=0.15$--$0.20$.
This outcome arises because the amount of $\mathrm{Ba}$ substitution required to achieve the tetragonal $I4/mmm$($\alpha = 180^\circ$) structure is sufficiently small, thereby the detrimental impact of hole doping on pairing interaction is minimized.

\section*{acknowledgments}
This work was supported by JST K Program Grant No. JPMJKP25Z3, JSPS KAKENHI Grant No. JP24K01333, JP22K04907, JP25K00959, JP26K08179 (K.K), JP25K08459, 26K01219 (H.S.), and JST-FOREST JPMJFR246T (H.S.).
The computing resource is supported by the supercomputer system (system-B) in the Institute for Solid State Physics, the University of Tokyo, and the supercomputer of Academic Center for Computing and Media Studies (ACCMS), Kyoto University.

\appendix*

\section{Formalism of multi-orbital FLEX}

We briefly summarize the formalism of multi-orbital FLEX adopted in this study.
We consider a multi-orbital Hamiltonian given in the present form,
\begin{eqnarray}
H_{\mathrm{int}}=&&U\sum_{i,\mu}n_{i\mu\uparrow}n_{i\mu\downarrow}
+U'\sum_{i,\mu<\nu,\sigma}n_{i\mu\sigma}n_{i\nu\bar{\sigma}}\nonumber\\ 
&&+(U'-J)\sum_{i,\mu<\nu,\sigma}n_{i\mu\sigma}n_{i\nu\sigma}\nonumber\\
&&-J\sum_{i,\mu\neq\nu}c^{\dagger}_{i\mu\uparrow}c_{i\mu\downarrow}c^{\dagger}_{i\nu\downarrow}c_{i\nu\uparrow}\nonumber\\
&&+J'\sum_{i,\mu\neq\nu}c^{\dagger}_{i\mu\uparrow}c^{\dagger}_{i\mu\downarrow}c_{i\nu\downarrow}c_{i\nu\uparrow},
\end{eqnarray}
where $U$, $U'$, $J$, and $J'$ are the intraorbital repulsion, the interorbital repulsion, Hund's coupling, and the pair hopping, respectively. 
Here, $i$ denotes the sites, $\mu$, $\nu$ the orbitals, and $\sigma$ the spins.
$c_{i\mu\sigma}$ and $c^{\dagger}_{i\mu\sigma}$ are annihilation and creation operators of electrons, respectively. 
The number operator is defined as $n_{i\mu\sigma}=c^{\dagger}_{i\mu\sigma}c_{i\mu\sigma}$.

In the FLEX approximation, the self energy is calculated by taking into account the bubble and ladder type diagrams that comprise the irreducible susceptibility 
\begin{equation}
\chi^{0}_{l_{1}l_{2}l_{3}l_{4}}(q)=-\frac{T}{N}\sum_{k}G_{l_{1}l_{3}}(k+q)G_{l_{4}l_{2}}(k),
\end{equation}
which is calculated from the renormalized Green function $G_{l_il_j}(k)$, where $k,q$ stands for both the wave vector and the Matsubara frequency, and $l_{i}$ denotes the orbitals. 
The Dyson equation is numerically solved to obtain the renormalized Green function in a self-consistent manner.
Using the converged Green function and the spin and charge susceptibilities,  
\begin{equation}
\hat{\chi}^{s}(q)=\frac{\hat{\chi}^{0}(q)}{1-\hat{S}\hat{\chi}^{0}(q)},
\end{equation}
\begin{equation}
\hat{\chi}^{c}(q)=\frac{\hat{\chi}^{0}(q)}{1+\hat{C}\hat{\chi}^{0}(q)},
\end{equation}
respectively, with the interaction matrices
\begin{equation}
S_{l_{1}l_{2}l_{3}l_{4}}=\left\{\begin{array}{cc}
U, & l_{1}=l_{2}=l_{3}=l_{4},\\
U', & l_{1}=l_{3}\neq l_{2}=l_{4},\\
J, & l_{1}=l_{4}\neq l_{2}=l_{3},\\
J', & l_{1}=l_{2}\neq l_{3}=l_{4},
\end{array}\right.
\end{equation}
\begin{equation}
C_{l_{1}l_{2}l_{3}l_{4}}=\left\{\begin{array}{cc}
U, & l_{1}=l_{2}=l_{3}=l_{4},\\
-U'+2J, & l_{1}=l_{3}\neq l_{2}=l_{4},\\
2U'-J, & l_{1}=l_{4}\neq l_{2}=l_{3},\\
J', & l_{1}=l_{2}\neq l_{3}=l_{4},
\end{array}\right.
\end{equation}
are plugged into the linearized Eliashberg equation
\begin{eqnarray}
  &&\lambda\Delta_{\mu\nu}(k)\nonumber\\
  &&=-\frac{T}{N}\sum_{q,m_1,m_2,m_3,m_4}V_{{\rm pair},\mu m_{1}m_{4}\nu}(q)G_{m_{1}m_{2}}(k-q)\nonumber\\
  &&\times \Delta_{m_{2}m_{3}}(k-q)G_{m_{4}m_{3}}(q-k)
\end{eqnarray}
to obtain eigenvalue $\lambda$ and eigenfunction $\Delta(k)$, where the effective spin-singlet pairing interaction is
\begin{equation}
V_{\rm pair}(q)=\frac{3}{2}\hat{S}\hat{\chi}^{s}(q)\hat{S}-\frac{1}{2}\hat{C}\hat{\chi}^{c}(q)\hat{C}+\frac{3}{4}\hat{S}+\frac{1}{4}\hat{C}.
\end{equation}

\bibliography{laba}

\begin{thebibliography}{91}
\bibitem [{com()}]{commentact}%
  \BibitemOpen
  \href@noop {} {}\bibinfo {note} {To our knowledge, their synthesis has not
  yet been experimentally realized}\BibitemShut {NoStop}%
\bibitem [{kco()}]{kconv}%
  \BibitemOpen
  \href@noop {} {}\bibinfo {note} {We have checked more dense ${\bm k}$-mesh up
  to a 12 $\times$ 12 $\times$ 12 ${\bm k}$-mesh. However, there is no
  significant differences with the 8 $\times$ 8 $\times$ 8 ${\bm k}$-mesh case,
  e.g., the difference of the lattice constant $c$ is about 0.01 \AA~~between
  the two cases. Therefore, we present the results of the 8 $\times$ 8 $\times$
  8 ${\bm k}$-mesh case in the main text.}\BibitemShut {Stop}%
\end{thebibliography}

\end{document}